\documentclass[twoside,fleqn]{article}
\usepackage{espcrc2}
\usepackage{amsmath,amssymb}

\def\BR{\operatorname{BR}}

\title{The rare decays of $D$ mesons}

\author{
S.\ Fajfer\address[FMF]{Physics Department, University of Ljubljana,
SI-1001 Ljubljana, Slovenia}\address[IJS]{J.\ Stefan Institute,
SI-1000 Ljubljana, Slovenia}\thanks{Talk given by S.\ Fajfer},
A.\ Prapotnik\addressmark[IJS],
S.\ Prelov\v sek\addressmark[FMF]\addressmark[IJS],
P.\ Singer\address[TEC]{Department of Physics, Technion -- Israel
Institute  of Technology, Haifa IL-32000, Israel}, and
J.\ Zupan\addressmark[IJS]
}

\begin{document}

\begin{abstract}
The flavor changing transitions in the $c \to u \gamma$, $c \to u
\gamma \gamma$ and $c \to u l^+ l^-$ offer the possibility to search
for new physics in the charm sector. We investigate dominant decay
mechanisms in the radiative decays $D \to V \gamma$, $D\to P (V) l^+
l^-$, $D \to \gamma \gamma$ and we discuss chances to see physics
beyond the standard model in these decays.  In addition, we analyze
Cabibbo allowed $D \to K \pi \gamma$ decays with nonresonant 
$K \pi$, and we probe the role of light vector mesons in these decays.
\end{abstract}

\maketitle

The search for physics beyond the standard model has been focused on
the down-like quark sector, while the up-like quark sector has been
less researched.  The rare $D$ decays offer an opportunity to
investigate the FCNC effects in the charm sector.  To date, no
radiative or dilepton weak decay of $D$ has been detected.  Only upper
bounds have been established so far for a sizable number of the
radiative or dilepton weak decay of $D$ mesons. The radiative decays
$D^0 \to \rho^0, $ $\omega^0,$ $ \phi, $ $ \bar K^{*0}$ $+ \gamma$
were recently bounded \cite{CLEO1} to branching ratios in the
$10^{-4}$ range, which is approaching the standard model expectations
(see, e.g. \cite{FPS2} where additional previous works are
listed). The dilepton decays $D \to P l^+ l^-$, $D \to V l^+ l^-$ are
the subject of intensive searches at CLEO and Fermilab
\cite{CLEO2} - \cite{E791}. Here again, with upper bounds of $\lesssim
10^{-5}$ for branching ratios of the various modes one approaches the
expectations of the standard model \cite{FPS3,BGHP,BGHP0,FPS1}. The
situation should improve in the future, due to new possibilities for
observation of charm meson decays at BELLE, BABAR and
Tevatron. Recently, upper limits in the $10^{-5} - 10^{-4}$ range were
established \cite{FOCUS,E791} also for $D^0$ dilepton decays with two
nonresonant pseudoscalar mesons in the final state $D^0 \to(\pi^+
\pi^-, K^- \pi^+, K^+ K^-)$ $ \mu^+ \mu^-$, though no comparable
results are available yet for similar photonic decays.
 
The inclusive $c \to u \gamma$ transition is strongly GIM suppressed
at one loop electroweak order giving the branching ratio of the order
$10^{-18}$.  In our approach we include the $c \to u \gamma$ short
distance contribution by using the Lagrangian
\begin{equation}
{\cal L}=-A\,C_{7\gamma}^{\text{eff}}\frac{e}{4\pi^2} F_{\mu \nu}
\big[\bar u \sigma^{\mu\nu} \tfrac{1}{2}(1 + \gamma_5)c\big],
\label{eq-103}
\end{equation}
where $m_c$ is a charm quark mass and $A = ( G_F/\sqrt{2}) \;V_{us}
V_{cs}^*$.  We follow \cite{GMW} and we take $C_{7\gamma}^{\text{eff}}=
(-0.7+ 2 i) \times 10^{-2}$.  The branching ratio induced by this QCD
corrected effective Lagrangian is $\BR(c \to u \gamma) \simeq 3
\times 10^{-8}$. A variety of models beyond the standard model were
investigated and it was found that the gluino exchange diagrams
\cite{PW} within general MSSM give the largest enhancement
\begin{equation}
\frac{ \BR(c \to u \gamma)_{\text{MSSM}}}{ \BR(c \to u \gamma)_{\text{SM}}} 
\simeq 10^2.
\label{1}
\end{equation}
Unfortunately, the long distance physics screens such effects since
it usually dominates the decay amplitude \cite{FPS2}.  The long
distance contribution is induced by the effective nonleptonic $
|\Delta c|=1 $ weak Lagrangian
\begin{align}
{\cal L} = &-A_{ij} \big[a_1\bar u\gamma^{\mu}(1-\gamma_5)q_i
\bar q_j\gamma_{\mu}(1-\gamma_5)c
\nonumber\\
&+a_2\bar q_j\gamma_{\mu}(1-\gamma_5)q_i\bar
u\gamma^{\mu}(1-\gamma_5)c\big],
\label{eff}
\end{align}
with $A_{ij} =( G_F/\sqrt{2} )\;V_{cq_j}^*V_{uq_i}$, accompanied by
the emission of the virtual photon. Here $q_{i,j}$ denote the $d$ or
$s$ quark fields. The effective  Wilson coefficients are 
$a_1=1.2$ and $a_2=-0.5$
\cite{FPS2}. In our calculations of the long distance effects we use
\cite{FPS2,FPS3,FPS1} the theoretical framework of heavy meson chiral
Lagrangian \cite{wise}. In the treatment of the $D$ mesons rare 
decays we use the factorization approximation for the calculation of
weak transition elements. We consider the use of this approach to be
justified by the "near" success of the approach for the nonleptonic
amplitudes.  In Table 1 we present the branching ratios of $D \to V
\gamma$ decays \cite{FPS2}. The uncertainty is due to relative unknown
phases of various contributions.
\begin{table}[h]
\begin{center}
\label{tab1}
\caption{ The branching ratios for $D \to V \gamma$ decays.}
\begin{tabular}{ll}
\hline
$D\to V \gamma$ & $\BR$ \\
\hline
$ D^0 \to {\bar K}^{*0} \gamma$ &$ [6- 36 ]\times 10^{-5}$ \\
$ D_s^+ \to \rho^+ \gamma$ &$ [20-80]\times 10^{-5} $ \\
$ D^0 \to \rho^{0} \gamma$&$ [0.1-1] \times 10^{-5}$ \\
$ D^0 \to \omega \gamma$ &$[ 0.1 - 0.9]\times 10^{-5}$  \\  
$ D^0 \to \Phi \gamma$ &$ [0.4 - 1.9 ]\times 10^{-5} $ \\
$ D^+ \to \rho^+ \gamma$ &$ [0.4 -6.3]\times 10^{-5}$\\
$ D_s^+ \to K^{*+ }\gamma$ &$[1.2 - 5.1]\times 10^{-5}$ \\
$ D^+ \to K^{*+} \gamma$ &$ [0.3- 4.4]\times 10^{-6}$ \\
$ D^0 \to K^{*0} \gamma$ &$ [0.3 - 2.0] \times 10^{-6}$   \\ 
\hline
\end{tabular}
\end{center}
\end{table}
Although the branching ratios are dominated by the long distance
contributions, the size of the short distance contribution can be
obtained from the difference of the decay widths $\Gamma (D^0
\to\rho^{0} \gamma)$ and $\Gamma(D^0 \to \omega \gamma)$ \cite{FPSW}.
Namely, the long distance mechanism $c\bar u \to d \bar d \gamma$
overshadows the $c\bar u \to u \bar u \gamma$ transition in $D^0 \to
\rho^{0} \gamma$ and $D^0 \to \omega \gamma$, the $ \rho^{0} $
and $\omega$ mesons being mixture of $u \bar u$ and $d \bar d$.
However, the LD contributions are mostly canceled in the ratio
\begin{align}
R &= \frac{ \BR(D^0 \to\rho^{0} \gamma) - \BR(D^0 \to \omega \gamma)}
{\BR (D^0 \to \omega \gamma)}
\nonumber\\
&\propto \operatorname{Re} \frac{ A(D^0 \to u \bar u \gamma)}{A(D^0
\to d \bar d \gamma)},
\label{2}
\end{align}
which is proportional to the SD amplitude $A(D^0 \to u \bar u \gamma)$
driven by $c \to u \gamma$.  The ratio $R$ is  $6 \pm 15 \%$ in the
standard model \cite{FPSW}, and can be enhanced up to ${\cal O}(1)$ in the
MSSM. In addition to the $c \to u \gamma$ searches in the charm meson
decays, we have suggested to search for this transition in $B_c \to
B_u^* \gamma$ decay \cite{FPS-B}, where the long distance contribution
is much smaller.

The $c\to ul^+l^-$ amplitude is given by the $\gamma$ and $Z$ penguin
diagrams and $W$ box diagram at one-loop electroweak order in the
standard model, and is dominated by the light quark contributions in
the loop.  In Table 2 we give the $c\to ul^+l^-$ branching ratios
calculated in the standard model and MSSM \cite{FPS1}.
\begin{table}[h]
\begin{center}
\label{tab2}
\caption{The $c\to ul^+l^-$ branching ratios.}
\begin{tabular}{lll}
\hline
$ \enspace $& $\BR^{\text{SM}}$ & $\BR^{\text{MSSM}}$\\
\hline
 $ c \to u e^+ e^-$ & 
$[6\pm1]\times 10^{-9}$&$6.0\times 10^{-8}$\\
 $ c \to u \mu^+ \mu^-$& $[6\pm1]\times 10^{-9}$ &$2.0\times
10^{-8}$\\
\hline
\end{tabular}
\end{center}
\end{table} 
The amplitudes for exclusive decays $D \to V l^+ l^-$ and $D \to P l^+
l^-$ are dominated by the long distance contributions.  The branching
ratios for these decays, as obtained in \cite{FPS3,FPS1}, are given in
Tables 3 and 4. In these Tables the first column represents the SD
contributions, while in the second column the rates coming from
the LD contributions are given. The rates for $ D \to V e^+e^-$ are comparable
to those in Table 3 and can be found in \cite{FPS3}.
\begin{table}[h]
\begin{center}
\label{tab3}
\caption{The $D\to V\mu^+ \mu^-$ branching ratios.}
\begin{tabular}{lll}
\hline
$ D\to V\mu^+ \mu^- $& $\BR_{\text{SD}}$ & $\BR_{\text{LD}}$\\
\hline
 $ D^0\to \bar K^{*0}\mu^+ \mu^- $ & $0$& 
$[1.6 - 1.9]\times 10^{-6}$\\
 $ D^+_s\to \rho^+\mu^+ \mu^- $ & $0$& 
$[3.0- 3.3]\times 10^{-5}$\\
 $ D^0\to \rho^{0}\mu^+ \mu^- $ & $9.7 \times 10^{-10}$& 
$[3.5-4.7]\times 10^{-7}$\\
 $ D^0\to \omega\mu^+ \mu^- $ & $9.1 \times 10^{-10}$& 
$[3.3-4.5]\times 10^{-7}$\\
 $ D^0\to \phi \mu^+ \mu^- $ & $0$& 
$[6.5-9.0] \times 10^{-8}$\\
 $ D^+\to \rho^+ \mu^+ \mu^- $ & $4.8 \times 10^{-9}$& 
[$1.5-1.8] \times 10^{-6}$\\
 $ D^+_s\to K^{*+} \mu^+ \mu^- $ & $1.6 \times 10^{-9}$& 
$ [5.0 -7.0] \times 10^{-7}$\\
 $ D^+\to K^{*+} \mu^+ \mu^- $ & $ 0$& 
$ [3.1-3.7]\times 10^{-8}$\\
$ D^0\to \ K^{*0}\mu^+ \mu^- $ & $0$& 
$[4.4-5.1] \times 10^{-9}$\\
\hline
\end{tabular}
\end{center}
\end{table}
\begin{table}[h]
\begin{center}
\label{tab4}
\caption{The $D\to P l^+ l^-$ ($l= e, \mu$) branching ratios.}
\begin{tabular}{lll}
\hline
$ D\to P l^+ l^-$& $Br_{SD}$ & $Br_{LD}$\\
\hline
 $ D^0\to \bar K^{0}l^+ l^-$ & $0$& 
$4.3\times 10^{-7}$\\
 $ D^+_s\to \pi^+l^+ l^-$ & $0$& 
$ 6.1\times 10^{-6}$\\
 $ D^0\to \pi^{0}l^+ l^- $ & $1.9 \times 10^{-9}$& 
$2.1 \times 10^{-7}$\\
 $ D^0\to \eta l^+ l^-$ & $2.5 \times 10^{-10}$& 
$4.9\times 10^{-8}$\\
 $ D^0\to \eta' l^+ l^-$ & $ 9.7 \times 10^{-12} $& 
$2.4 \times 10^{-8}$\\
 $ D^+\to \pi^+ l^+ l^-$ & $9.4\times 10^{-9}$& 
$1.0 \times 10^{-6}$\\
 $ D^+_s\to K^{+} l^+ l^-$ & $ 9.0\times 10^{-10}$& 
$ 4.3 \times 10^{-8}$\\
$ D^+\to K^{+} l^+ l^-$ & $ 0$& 
$ 7.1 \times 10^{-9}$\\
$ D^0\to \ K^{0}l^+ l^- $ & $0$& 
$1.1 \times 10^{-9}$\\
\hline
\end{tabular}
\end{center}
\end{table}
There is a significant improvement of the experimental upper bounds
for $D \to P l^+l^-$ decay rates, as recently obtained by FOCUS \cite{FOCUS},
with new upper bounds of $10^{-5}$ or less, close to the theoretical
predictions \cite{BGHP0,FPS1}.  The allowed kinematic region for the
dilepton mass $m_{ll}$ in the $D\to Pl^+l^-$ decay is
$m_{ll}=[2m_l,m_D-m_P]$. The long distance contribution has resonant
shape with poles at $m_{ll}=m_{\rho^0},~m_\omega,~m_\phi$
\cite{BGHP0,FPS1}. There is no pole at $m_{ll}=0$ since the decay
$D\to P\gamma$ is forbidden. The short distance contribution is rather
flat.  The spectra
 of $D\to P e^+e^-$ and $D\to P\mu^+\mu^-$ decays
in terms of $m_{ll}$ are practically identical. The difference in
their rates due to the kinematic region $m_{ll}=[2m_e,2m_\mu]$ is
small and we do not consider them separately.

The differential distribution for $D^{+,0} \to \pi^{+0} l^+ l^-$
\cite{BGHP0,FPS1} indicates that the high mass dilepton region might
give an opportunity for detecting $c \to u l^+ l^-$. Before making a
definite statement on such possibility, we should examine this
kinematical region of high dilepton mass in $D\to \pi l^+l^-$ decays
more closely. For instance, in this region the excited states of the
vector mesons $\rho$, $\omega$ and $\phi$ may become important. 
Even
with the presence of these states we found that the only viable
channel for investigating the $c \to u l^+ l^-$ transition is $D \to
\pi l^+ l^-$, in the kinematic region of the dilepton invariant mass
$m_{ll}$ above the resonance $\phi$, where the long distance
contribution is reduced. The kinematics of the processes $D \to V l^+
l^-$ would be more favorable to probe the possible supersymmetric
enhancement at small $m_{ll}$, but the long distance contributions in
these channels are even more disturbing \cite{FPS3}.

Motivated by the experimental efforts to observe rare D meson decays
\cite{Selen}, and noticing that $B_s \to \gamma \gamma$ offers
possibility to observe physics beyond the SM, we undertook an
investigation of the $D^0\to \gamma \gamma$ decay \cite{FSZ}.  The
short distance contribution is expected to be rather small, as already
encountered in the one photon decays \cite{BGHP}, hence the main
contribution would come from long distance interactions.  The total
amplitude is dominated by terms proportional to $a_1$ that contribute
only through loops with Goldstone bosons. Loop contributions
proportional to $a_2$ vanish at this order.  We remark that the
contribution of the order ${\cal O}(p)$ does not exist in the $D^0\to
\gamma \gamma$ decay, and the amplitude starts with contribution of
the order ${\cal O}(p^3)$. The chiral loops of order ${\cal O}(p^3)$
are finite, as they are in the similar case of $K\to \gamma \gamma$
decays. At this order the amplitude receives also an annihilation type
contribution proportional to the $a_2$ Wilson coefficient, given by
the Wess-Zumino anomalous term coupling light pseudoscalars to two
photons.  Terms which contain the anomalous electromagnetic coupling
of the heavy quark Lagrangian are suppressed compared to the leading
loop effects \cite{FSZ}. The invariant amplitude for $D^0 \to \gamma
\gamma $ decay can be written using gauge and Lorentz invariance in
the following form,
\begin{align}
M = \Big[ i &M^{(-)} \big(g^{\mu \nu} -\frac{k_2^\mu k_1^\nu}{k_1
\cdot k_2} \big)
\nonumber\\
+ &M^{(+)} \epsilon^{\mu \nu\alpha\beta}k_{1\alpha}k_{2\beta}\Big]
\epsilon_{1\mu}\epsilon_{2\nu},
\label{eq-104}
\end{align}
where $M^{(-)}$ is a parity violating and $M^{(+)}$ a parity
conserving part of the amplitude, while $k_{1(2)}$, $\epsilon_{1(2)}$
are respectively the four momenta and the polarization vectors of the
outgoing photons.  We give in Table 5 the numerical results for the
amplitudes originating from the different contributions.
\begin{table} [h]
\begin{center}
\label{Table5}
\caption{The nonvanishing amplitudes in  $D^0 \to \gamma \gamma $. The
amplitudes coming from the anomalous and short distance
$C_{7\gamma}^{\text{eff}}$ Lagrangians are given in first two lines. The
finite and gauge invariant sums of one-loop amplitudes are listed in
the next three lines ($M_i^{(\pm)}=\sum_j M_{i.j}^{(\pm)}$).  The
numbers $1,2,3$ denote the row of diagrams on the Fig. 2 in
\cite{FSZ}.  In the last line the sum of all amplitudes is given. All
amplitudes are given in the units $10 ^{-10}$ ${\rm \;GeV}$.}
\begin{tabular}{lllll}
\hline
&$M_{ i}^{(-)}$&& $M_{i}^{(+)}$&
\\
\hline
A & $0$ & & $-0.53$
\\
SD & $-0.27$&$-0.81 i$ & $-0.16$&$ -0.47 i$
\\
$1$ & $3.55$&$+9.36i$ & $0$
\\
$2$ &$1.67$ & &  $0$
\\
$3$ & $-0.54$&$+2.84i$&$0$
\\
\hline
$\sum_i  M_i^{(\pm)}$& $4.41$&$+11.39 i$ &$-0.69$&$ -0.47 i$
\\
\hline
\end{tabular}
\end{center}
\end{table}
Within this framework, the leading contributions are found to arise
from the charged $\pi$ and $K$ mesons running in the chiral loops and
our calculation predicts that the $D\to 2 \gamma$ decay is mostly a
parity violating transition \cite{FSZ}. We estimate that the total
uncertainty is not larger than 50\% \cite{FSZ}, including possible
effects from light resonances like $\rho$, $K^*$, $a_0(980)$,
$f_0(975)$. Accordingly, we conclude that the predicted branching
ratio is
\begin{equation}
\BR(D^0\to \gamma \gamma)= (1.0 \pm 0.5)\times 10^{-8}. 
\label{fin-res}
\end{equation}

We look forward to experimental attempts of detecting this decay. Our
result suggests that the observation of $D\to 2 \gamma$ at a rate
which is an order of magnitude larger than (\ref{fin-res}), could be a
signal for the type of "new physics" \cite{PW,FSZ}.

In addition to the $D$ meson rare decays in which the FCNC transitions
might occur, we undertook a study of the Cabibbo allowed radiative
decays $D^+ \to {\bar K}^0 \pi^+ \gamma$ and $D^0 \to K^- \pi^+
\gamma$ with nonresonant $K$ $\pi$, which we consider to be the most
likely candidates for early detection. Here we used the heavy quark
chiral Lagrangian supplemented by light vector mesons, as the
theoretical framework.

These decays are the charm sector counterpart of the $K \to \pi \pi
\gamma$ decays, which have provided a wealth of information on meson
dynamics.  Using the factorization approximation for the calculation
of weak transition elements we use the information obtained from
semileptonic decays \cite{castwo,BFO2}.  The nonleptonic $ D \to K\pi$
amplitude cannot be calculated accurately in the factorization
approximation from the diagrams provided by our model. Such a
calculation gives a rather good result for the $D^+ \to {\bar K}^0
\pi^+$ channel but is less successful for the $D^0 \to K^+ \pi^-$
decay. In order to overcome this deficiency and to be able to present
accurately the bremsstrahlung component of the radiative transition,
we shall use an alternative approach for its derivation. 
It means, we  take  the experimental values for the 
$D \to K \pi$ amplitudes, in the calculation of
the bremsstrahlung component. In order to accommodate this, we write
the decay amplitude as
\begin{align}
{\cal M} = &-\frac{G_f}{\sqrt{2}}V_{cs}V_{du}^*
\{F_0 [\frac{q \cdot\varepsilon}{q  \cdot k}-
\frac{p\cdot \varepsilon}{p\cdot k}]
\nonumber\\
&+F_1 [ (q \cdot\varepsilon) (p  \cdot k)- 
(p \cdot\varepsilon) (q  \cdot k)]
\nonumber\\
&+F_2 \varepsilon^{\mu \alpha \beta \gamma}
\varepsilon_\mu v_\alpha k_\beta q_\gamma \},
\label{amp1}
\end{align}
where $F_0$ is the experimentally determined $D \to K\pi$ amplitude
and $F_1$, $F_2$ are the form factors of the electric and magnetic
direct transitions which we calculate with our model.  When
intermediate states appear to be on the mass shell, we use Breit
Wigner formula.  Thus, we get for the branching ratios of the electric
transitions including bremsstrahlung, with $|F_0|$ determined
experimentally and taking the photon energy cut $E_c\geq 100\,$MeV
\begin{align}
\BR(D^+ &\to \bar K^0 \pi^+ \gamma)_{\text{PV,ex}}^{E_c}
\nonumber\\
&= (2.3-2.5)\times 10^{-4}. 
\label{rpv+2}
\end{align}
For the $D^0$ radiative decay we get
\begin{align}
\BR(D^0 &\to K^- \pi^+ \gamma)_{\text{PV,ex}}^{E_c}  
\nonumber\\
&= (4.3-6.0)\times 10^{-4}.
\label{rpv02}
\end{align}
The uncertainty in the $F_0/F_1$ phase is less of a problem in $D^+
\to \bar K^0 \pi^+ \gamma$ than in $D^0 \to K^- \pi^+ \gamma$. If we
take the bremsstrahlung amplitude alone as determined from the
knowledge of $|F_0|$, disregarding the direct electric $F_1$ term, the
above numbers are replaced by $2.3\times 10^{-4}$ for $D^+$ decay and
$5.5\times 10^{-4}$ for the $D^0$ decay. The contribution of the
direct parity violating term (putting $F_0 = 0$), is $\text{BR}(D^+ \to \bar
K^0 \pi^+ \gamma)_{\text{dir,PV} }=1.0 \times 10^{-5}$ and $\text{BR}(D^0 \to K^-
\pi^+ \gamma)_{\text{dir,PV}}= 1.64 \times 10^{-4}$. For the parity
conserving direct magnetic transition we get
\begin{align}
&\BR(D^+ \to \bar K^0 \pi^+ \gamma)_{\text{PC}} = 
2.0 \times 10^{-5}, 
\label{rpv+2m}
\\
&\BR(D^0 \to K^- \pi^+ \gamma)_{\text{PC}} = 
1.4\times 10^{-4}.
\label{rpv02m}
\end{align}
Hence, the two direct transitions are predicted to be of about the
same strength.

If we disregard the contribution of vector mesons to the direct part
of the radiative decays, the parity-conserving part of the amplitude
is considerably decreased, by one order of magnitude in the rate in
$D^+ \to \bar K^0 \pi^+ \gamma$ decay and by two orders of magnitude
in $D^0 \to K^- \pi^+ \gamma$. On the other hand, their contribution
is not felt in a significant way in the parity - violating part of
the amplitudes. In any case, the detection of the direct part of these
decays at the predicted rates, will constitute a proof of the
important role of the light vector mesons.

We conclude by expressing the hope that the interesting features which
these decays provide and were analyzed in \cite{FAS}, will bring to an
experimental search in the near future.

\end{document}